\documentclass[twocolumn,twocolappendix]{aastex631}
\pdfoutput=1
\usepackage{amsmath,amstext}
\usepackage[T1]{fontenc}
\usepackage{times}
\usepackage{apjfonts} 
\usepackage[figure,figure*]{hypcap}
\usepackage{natbib}
\usepackage{hyperref}
\usepackage{booktabs}
\usepackage{graphicx}
\usepackage{multirow}
\usepackage[encapsulated]{CJK}
\usepackage[caption=false]{subfig}
\graphicspath{{./}{Figures/}}
\newcommand{\grizli}{\textsc{Gri$z$li}}
\newcommand{\eazy}{\textsc{EA$z$Y}}
\newcommand{\pros}{\textsc{Prospector}}
\newcommand{\prosb}{\pros{}-$\beta$}
\newcommand{\galfit}{\textsc{GALFIT}}
\newcommand{\deltams}[1]{\Delta{\rm MS_{#1}}}

\received{2023 December 22}
\revised{2024 April 18}
\submitjournal{the Astrophysical Journal Letters}
\shorttitle{Two Distinct Classes of Quiescent Galaxies}
\shortauthors{Cutler et al.}

\begin{document}
\title{Two Distinct Classes of Quiescent Galaxies at Cosmic Noon Revealed by JWST PRIMER and UNCOVER}

\correspondingauthor{Sam E. Cutler}
\email{secutler@umass.edu}

\author[0000-0002-7031-2865]{Sam E. Cutler}
\affiliation{Department of Astronomy, University of Massachusetts, Amherst, MA 01003, USA}

\author[0000-0001-7160-3632]{Katherine E. Whitaker}
\affiliation{Department of Astronomy, University of Massachusetts, Amherst, MA 01003, USA}
\affiliation{Cosmic Dawn Center (DAWN), Denmark}

\author[0000-0003-1614-196X]{John R. Weaver}
\affiliation{Department of Astronomy, University of Massachusetts, Amherst, MA 01003, USA}

\author[0000-0001-9269-5046]{Bingjie Wang (\begin{CJK*}{UTF8}{gbsn}王冰洁\ignorespacesafterend\end{CJK*})}
\affiliation{Department of Astronomy \& Astrophysics, The Pennsylvania State University, University Park, PA 16802, USA}
\affiliation{Institute for Computational \& Data Sciences, The Pennsylvania State University, University Park, PA 16802, USA}
\affiliation{Institute for Gravitation and the Cosmos, The Pennsylvania State University, University Park, PA 16802, USA}

\author[0000-0002-9651-5716]{Richard Pan}
\affiliation{Department of Physics \& Astronomy, Tufts University, MA 02155, USA}

\author[0000-0001-5063-8254]{Rachel Bezanson}
\affiliation{Department of Physics and Astronomy and PITT PACC, University of Pittsburgh, Pittsburgh, PA 15260, USA}

\author[0000-0001-6278-032X]{Lukas J. Furtak}
\affiliation{Physics Department, Ben-Gurion University of the Negev, P.O. Box 653, Be’er-Sheva 84105, Israel}

\author[0000-0002-2057-5376]{Ivo Labbe}
\affiliation{Centre for Astrophysics and Supercomputing, Swinburne University of Technology, Melbourne, VIC 3122, Australia}

\author[0000-0001-6755-1315]{Joel Leja}
\affiliation{Department of Astronomy \& Astrophysics, The Pennsylvania State University, University Park, PA 16802, USA}
\affiliation{Institute for Computational \& Data Sciences, The Pennsylvania State University, University Park, PA 16802, USA}
\affiliation{Institute for Gravitation and the Cosmos, The Pennsylvania State University, University Park, PA 16802, USA}

\author[0000-0002-0108-4176]{Sedona H. Price}
\affiliation{Department of Physics and Astronomy and PITT PACC, University of Pittsburgh, Pittsburgh, PA 15260, USA}

\author[0000-0001-8551-071X]{Yingjie Cheng}
\affiliation{Department of Astronomy, University of Massachusetts, Amherst, MA 01003, USA}

\author[0009-0001-4005-5490]{Maike Clausen}
\affiliation{Max Planck Institute of Astronomy, Königstuhl 17, 69117 Heidelberg, Germany}

\author[0000-0002-3736-476X]{Fergus Cullen}
\affiliation{Institute for Astronomy, University of Edinburgh, Royal Observatory, Edinburgh, EH9 3HJ, UK}

\author[0000-0001-8460-1564]{Pratika Dayal}
\affiliation{Kapteyn Astronomical Institute, University of Groningen, P.O. Box 800, 9700 AV Groningen, The Netherlands}

\author[0000-0002-2380-9801]{Anna de Graaff}\affiliation{Max-Planck-Institut f\"ur Astronomie, K\"onigstuhl 17, D-69117, Heidelberg, Germany}

\author[0000-0001-5414-5131]{Mark Dickinson}
\affiliation{NSF's National Optical-Infrared Astronomy Research Laboratory, 950 N. Cherry Ave., Tucson, AZ 85719, USA}

\author[0000-0002-1404-5950]{James S. Dunlop}
\affiliation{Institute for Astronomy, University of Edinburgh, Royal Observatory, Edinburgh, EH9 3HJ, UK}

\author[0000-0002-1109-1919]{Robert Feldmann}
\affiliation{Institute for Computational Science, University of Zurich, Zurich CH-8057, Switzerland}

\author[0000-0002-8871-3026]{Marijn Franx}
\affiliation{Leiden Observatory, Leiden University, P.O. Box 9513, 2300 RA Leiden, The Netherlands}

\author[0000-0002-7831-8751]{Mauro Giavalisco}
\affiliation{Department of Astronomy, University of Massachusetts, Amherst, MA 01003, USA}

\author[0000-0002-3254-9044]{Karl Glazebrook}
\affiliation{Centre for Astrophysics and Supercomputing, Swinburne University of Technology, PO Box 218, Hawthorn, VIC 3122, Australia}

\author[0000-0002-5612-3427]{Jenny E. Greene}
\affiliation{Department of Astrophysical Sciences, Princeton University, 4 Ivy Lane, Princeton, NJ 08544}

\author[0000-0001-9440-8872]{Norman A. Grogin}
\affiliation{Space Telescope Science Institute, Baltimore, MD 21218, USA}

\author[0000-0002-8096-2837]{Garth Illingworth}
\affiliation{Department of Astronomy and Astrophysics, University of California, Santa Cruz, CA 95064, USA}

\author[0000-0002-6610-2048]{Anton M. Koekemoer}
\affiliation{Space Telescope Science Institute, Baltimore, MD 21218, USA}

\author[0000-0002-5588-9156]{Vasily Kokorev}
\affiliation{Kapteyn Astronomical Institute, University of Groningen, P.O. Box 800, 9700 AV Groningen, The Netherlands}

\author[0000-0001-9002-3502]{Danilo Marchesini}
\affiliation{Department of Physics \& Astronomy, Tufts University, MA 02155, USA}

\author[0000-0003-0695-4414]{Michael V. Maseda}
\affiliation{Department of Astronomy, University of Wisconsin-Madison, 475 N. Charter St., Madison, WI 53706 USA}

\author[0000-0001-8367-6265]{Tim B. Miller}
\affiliation{Center for Interdisciplinary Exploration and Research in Astrophysics (CIERA), Northwestern University, 1800 Sherman Ave, Evanston IL 60201, USA}

\author[0000-0003-2804-0648 ]{Themiya Nanayakkara}
\affiliation{Centre for Astrophysics and Supercomputing, Swinburne University of Technology, PO Box 218, Hawthorn, VIC 3122, Australia}

\author[0000-0002-7524-374X]{Erica J. Nelson}
\affiliation{Department for Astrophysical and Planetary Science, University of Colorado, Boulder, CO 80309, USA}

\author[0000-0003-4075-7393]{David J. Setton}
\altaffiliation{Brinson Prize Fellow}
\affiliation{Department of Astrophysical Sciences, Princeton University, 4 Ivy Lane, Princeton, NJ 08544, USA}

\author[0009-0007-1787-2306]{Heath Shipley}
\affiliation{Department of Physics, Texas State University, San Marcos, TX 78666, USA}

\author[0000-0002-1714-1905]{Katherine A. Suess}
\altaffiliation{NHFP Hubble Fellow}
\affiliation{Kavli Institute for Particle Astrophysics and Cosmology and Department of Physics, Stanford University, Stanford, CA 94305, USA}

\begin{abstract}\noindent
We present a measurement of the low-mass quiescent size-mass relation at Cosmic Noon ($1<z<3$) from the JWST PRIMER and UNCOVER treasury surveys, which highlights two distinct classes of quiescent galaxies. While the massive population is well studied at these redshifts, the low-mass end has been previously under-explored due to a lack of observing facilities with sufficient sensitivity and spatial resolution. We select a conservative sample of low-mass quiescent galaxy candidates using rest-frame $UVJ$ colors and specific star formation rate criteria and measure galaxy morphology in both rest-frame UV/optical wavelengths (F150W) and rest-frame near-infrared (F444W). We confirm an unambiguous flattening of the low-mass quiescent size-mass relation, which results from the separation of the quiescent galaxy sample into two distinct populations at $\log(M_\star/M_\odot)\sim10.3$: low-mass quiescent galaxies that are notably younger and have disky structures, and massive galaxies consistent with spheroidal morphologies and older median stellar ages. These separate populations imply mass quenching dominates at the massive end while other mechanisms, such as environmental or feedback-driven quenching, form the low-mass end. This stellar mass dependent slope of the quiescent size-mass relation could also indicate a shift from size growth due to star formation (low masses) to growth via mergers (massive galaxies). The transition mass between these two populations also corresponds with other dramatic changes and characteristic masses in several galaxy evolution scaling relations (e.g. star-formation efficiency, dust obscuration, and stellar-halo mass ratios), further highlighting the stark dichotomy between low-mass and massive galaxy formation.
\end{abstract}


\section{Introduction}
The study of galaxy evolution has long sought how to link a galaxy's size and morphology to its star-formation history, especially in terms of determining how and why galaxies quench. Previous studies find that quiescent galaxies are more compact than their star-forming counterparts at a given mass out to cosmic noon \citep[$z\gtrsim2$; e.g.,][]{Kriek2009,Williams2010,Wuyts2011,vanderWel2014,Whitaker2015,Nedkova2021}. At higher stellar masses ($\log(M_\star/M_\odot)>10.5$), quiescent galaxies sizes depend strongly on stellar mass \citep[e.g.,][]{Damjanov2009,vanDokkum2009,vanderWel2014,Mowla2019,Cutler2022,Ito2023}, possibly due to rapid growth of an outer envelope caused by dry mergers \citep{Bezanson2009,Naab2009,Trujillo2011,Patel2013,Ownsworth2014,vanDokkum2015}. Simultaneously, quenching in the massive regime is thought to be primarily caused by galaxies reaching a stellar/halo mass threshold \citep[i.e., mass quenching,][]{YPeng2010}, at which point future star formation is primarily thought to be halted by cosmological starvation \citep[e.g.,][]{Feldmann2015}, shock heating of circumgalactic gas \citep[e.g.,][]{Dekel2019}, or active galactic nuclei (AGN) feedback.

\begin{table*}[ht!]
    \centering
    \begin{tabular}{l|c|c|c||c}
         & UNCOVER & PRIMER-COSMOS & PRIMER-UDS & Total  \\\hline\hline
        Total Photometric Catalog & 59845 & 58004 & 75914 & 193763\\
        $1<z<3$ and $7<\log(M_\star/M_\odot)<11$ & 4432 & 4610 & 6934 & 15976\\\hline
        $UVJ$ Quiescent ($C_Q>0.49$ \& $t_{50}>30$ Myr) & 218 & 145 & 266 & 629\\
        Quiescent ($UVJ$ Quiescent \& $\deltams{10}<-0.5$ \& ${\rm SFR}_{10}-{\rm SFR}_{100}<0.3$) & 108 & 99 & 147 & 354\\
        \textbf{Quiescent with \eazy{} Color Outliers Removed} & \textbf{94} & \textbf{96} & \textbf{143} & \textbf{333}\\\hline
        Quiescent ($t_{50}>500$ Myr) & 39 & 69 & 101 & 209\\
        Quiescent ($100<t_{50}<500$ Myr) & 27 & 20 & 25 & 72\\
        Quenched ($30<t_{50}<100$ Myr) & 28 & 7 & 17 & 52\\        
    \end{tabular}
    \caption{Sample size of the quiescent galaxy sample and sub-samples. The final adopted sample of low-mass $1<z<3$ galaxies is shown in bold.}
    \label{tab:samplesize}
\end{table*}

At lower stellar masses ($\log(M_\star/M_\odot)<10$), several recent studies find that the quiescent size-mass relation flattens, with sizes becoming comparable to star-forming galaxies of similar mass \citep{Dutton2011,Lange2015,Whitaker2017,Mowla2019,Nedkova2021,Kawinwanichakij2021,Cutler2022,Yoon2023}. These galaxies are too low-mass to be quenched by mass-related mechanisms: their halos are below the $\log(M_{\rm halo}/M_\odot)>11.8$ limit for shock heating the circumgalactic medium \citep{Dekel2019} and AGN feedback would only temporarily remove gas from the system before it can replenish \citep{Tacchella2016,Greene2020}. One potential mechanism for quenching these low-mass systems is by rapidly consuming the available gas via central starbursts, either triggered by mergers \citep{Puglisi2019} or disk/gas instabilities \citep{Dekel2014,Zolotov2015}. However, several studies show that this quenching event is only temporary, with these galaxies returning to the main sequence after the starburst or temporary quenching event \citep{Tacchella2016,Cutler2023}, likely due to the resumption of cold gas accretion. Simulations also suggest that star formation in these low-mass ($7<\log(M_\star/M_\odot)<9$) galaxies is bursty at higher redshifts \citep[e.g.,][]{Angles-Alcazar2017,Faucher-Giguere2018,Ma2018,Sun2023,Dome2023}, further evidence supporting that galaxies that appear quiescent at the epoch of observation may only be temporarily so. These ``mini-quenched'' galaxies \citep[e.g.,][]{Dome2023,Looser2023b,Strait2023} are likely the result of the interplay between gas inflow (via infalling cold gas streams or mergers) and gas removal (via stellar feedback or environmental interactions) and may be contaminants in finding a population of ``fully quenched'' low-mass galaxies.

Measurements of galaxy sizes can provide insight into how galaxies form, grow in size, and cease star formation. Previous measurements of the low-mass quiescent size-mass relation are limited by the spatial resolution and depth of observations possible at the time. For example, the Hubble Space Telescope (HST) WFC3 F160W has a half width at half max (HWHM) of 0.65 kpc at $z=2$, roughly the size at which the quiescent size-mass relation reaches a minimum and begins to flatten \citep[e.g.,][]{Nedkova2021,Cutler2022}. Likewise, robust size measurements have only been available to $\log(M_\star/M_\odot)>9.5$ at $z\sim1.5$ and $\log(M_\star/M_\odot)>10.5$ at $z\sim2.5$ with deep HST data \citep[e.g., Hubble Frontier Fields, CANDELS, and COSMOS,][]{Nedkova2021,Cutler2022}. The wavelength coverage of HST ($\lambda_{\rm rest}\leq0.8~\micron$ at $z=1$) also means observed sizes may have biased mass-to-light ratios due to ``outshining'' \citep{Papovich2001}, in which young stars dominate the light from a galaxy such that a significant fraction of the stellar mass can be missed with rest-optical data. Outshining causes light-weighted sizes of quiescent galaxies to be significantly larger than mass-weighted sizes at $\log(M_\star/M_\odot)\gtrsim9.5$, which could artificially lead to larger low-mass sizes should these trends hold \citep{Suess2019,vanderWel2023}.

New observations from the James Webb Space Telescope (JWST) NIRCam instrument provide significant improvements in depth, spatial resolution, and redder wavelength coverage over HST \citep{Rieke2023,Rigby2023}. Several studies have already leveraged these advancements to explore galaxy sizes out to $z\sim8$ \citep{Ormerod2023}, examine the evolution of the star-forming size-mass relation since $z=5.5$ \citep{Ward2023}, and investigate the dependence of S\'ersic-based measurements on wavelength \citep{Martorano2023}.

In this letter, we measure the sizes of 333 quiescent galaxies from JWST Ultradeep NIRSpec and NIRCam ObserVations before the Epoch of Reionization \citep[UNCOVER,][]{Bezanson2022} and Public Release IMaging for Extragalactic Research \citep[PRIMER,][]{Dunlop2021} in F150W and F444W, using 2D S\'ersic fits from \galfit{} \citep{Peng2002,Peng2010}. We assume a \cite{Chabrier2003} initial mass function and WMAP9 cosmology: $H_0=69.32~{\rm km~s^{-1}~Mpc^{-1}}$, $\Omega_M=0.2865$, and $\Omega_\Lambda=0.7135$ \citep{Hinshaw2013}.

\begin{figure*}[t!]
    \centering
    \includegraphics[width=\linewidth]{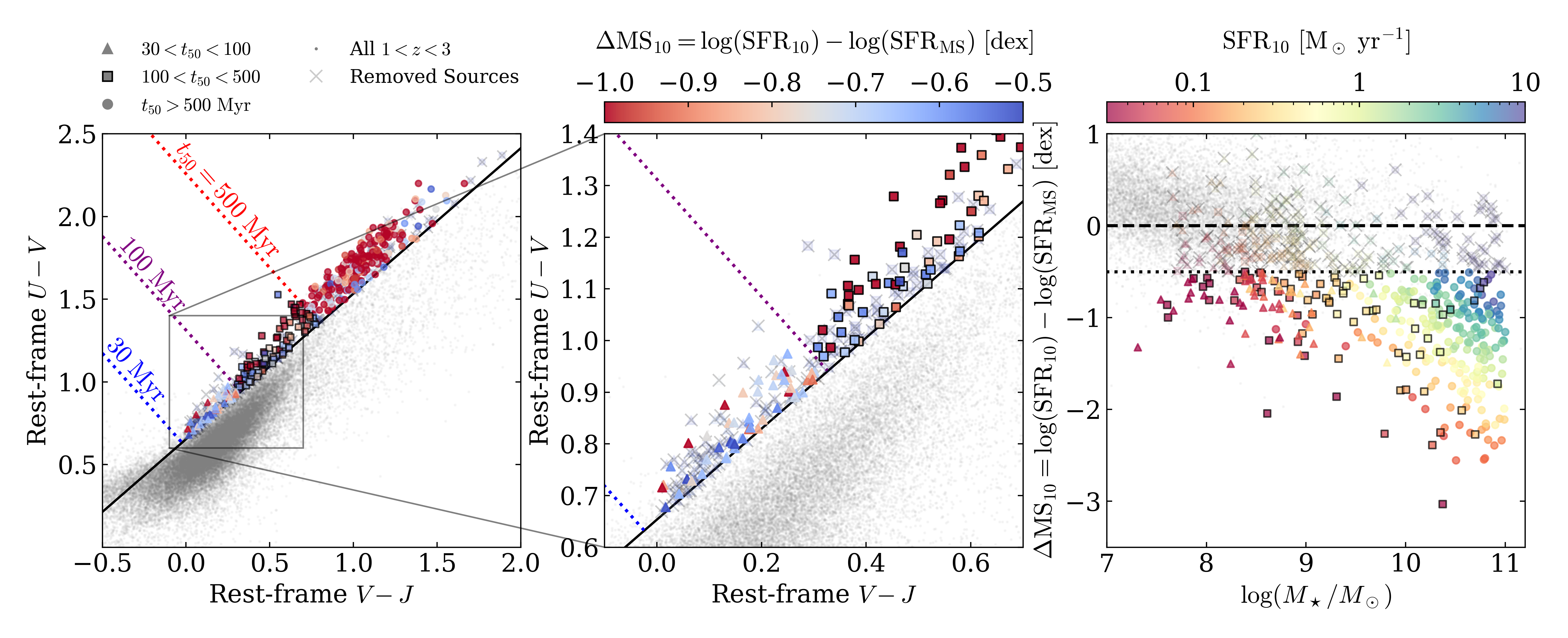}
    \caption{The rest-frame colors and star-formation properties of the 333 quiescent galaxies at $1<z<3$ with $\log(M_\star/M_\odot)<11$, as defined in Section \ref{sec:sample}. Sources are identified as $t_{50}>500$ Myr quiescent (circles), 100-500 Myr quiescent (black-outlined squares), or 30-100 Myr quenched galaxies (triangles), respectively, relative to the parent sample at $1<z<3$ (small grey points). The left panel shows rest-frame $UVJ$ colors from best-fit SPS models, with an inset (center) to show more detail, colored by $\deltams{10}$ (Eqn. \ref{eqn:dms}). Solid black and dotted lines show the $UVJ$ selections for general quiescent galaxies and cuts selecting galaxies of varying median age, respectively. The degree of quiescence for the sample is captured by difference in SFR relative to the average SFMS (dashed line), $\deltams{10}$, colored by the SFR at 10 Myr (right). Galaxies 0.5 dex below the SFMS (dotted line) are considered quenched. Sources marked with black x’s are potential contaminants and have been removed from the sample.} 
    \label{fig:sampleplots}
\end{figure*}

\section{Data and Sample Selection}\label{sec:sample}
\subsection{Imaging and Catalogs}
Our sample is built on the JWST UNCOVER and PRIMER surveys. UNCOVER (JWST-GO-2561) targets 45 sq. arcmin of the Abell-2744 lensing cluster with NIRCam F090W, F115W, F150W, F200W, F277W, F356W, F410M, and F444W and is the deepest-to-date (when augmented by strong lensing) publicly available survey \citep{Bezanson2022}. PRIMER (JWST-GO-1837) is a deep, wide-area survey that covers 378 sq. arcmin. of two HST legacy fields (COSMOS and UDS) with homogeneous depth in the same NIRCam bands as UNCOVER. PRIMER and UNCOVER are chosen because together they probe a wider range of luminosity and galaxy stellar mass than individually. The deeper imaging of UNCOVER finds more low-mass, faint galaxy populations, while the larger area of PRIMER enables us to detect more of the rarer, higher-mass/brighter galaxies. Moreover, the use of these surveys covers three separate fields, which reduces the impact of cosmic variance. Between $1<z<3$, UNCOVER reaches 95\% mass-completeness limits of $\log(M_\star/M_\odot)\sim7.3$ (7.8) at $z=1$ ($z=3$) for a sample of all galaxies, though older or quiescent galaxy samples may be less complete. In both surveys, samples are selected from a F277W-F356W-F444W detection image, with F444W down to $m_{5\sigma}=29.21$ ABmag in UNCOVER and 28.17 ABmag in PRIMER.

Imaging is from the v7\footnote{\url{https://dawn-cph.github.io/dja/imaging/v7/}} mosaics reduced by \grizli{} \citep{grizli} and rescaled to a 40 mas pixel scale in both F150W and F444W. The UNCOVER mosaics with bright cluster galaxy, intracluster light, and sky background subtraction are used \citep{Weaver2023}, while additional sky background is subtracted from the PRIMER science mosaics. Sample galaxies are selected from the UNCOVER \citep{Weaver2023} and PRIMER photometric catalogs as follows. The PRIMER catalogs (both COSMOS and UDS) are built using the \textsc{Aperpy}\footnote{\url{https://github.com/astrowhit/aperpy}} aperture photometry code, with settings identical to \cite{Weaver2023}. Both the UNCOVER and PRIMER catalogs are PSF-matched to F444W. The data presented in this article were obtained from the Mikulski Archive for Space Telescopes (MAST) at the Space Telescope Science Institute. The specific observations analyzed for UNCOVER and PRIMER can be accessed at \dataset[DOI: 10.17909/nftp-e621]{https://doi.org/10.17909/nftp-e621} and \dataset[DOI: 10.17909/ee2f-st77]{https://doi.org/10.17909/ee2f-st77}.

Redshifts and stellar populations properties for the UNCOVER~\footnote{The catalog and related documentation are accessible via the UNCOVER survey webpage (\url{https://jwst-uncover.github.io/DR2.html\#SPSCatalogs}) or Zenodo (\dataset[doi:10.5281/zenodo.8401181]{https://doi.org/10.5281/zenodo.8401181})} and the PRIMER photometric catalogs are inferred following \cite{Wang2023b}, using the \prosb{} model \citep{Wang2023} within the \pros{} \citep{prospector2021} Bayesian inference framework. \prosb{} provides robust photometric redshifts, rest-frame colors, and key stellar population properties, including stellar masses and non-parametric star-formation histories (SFHs). For the Abell 2744 field, we adopt the strong lens model from \cite{Furtak2023}. Note that the redshifts and stellar populations properties are inferred jointly, and that since the magnification factor depends on redshift, we account for lensing modification consistently during model fitting (see Section 3.1 in \citealt{Wang2023b} for details). This way, the scale-dependent priors (i.e., the mass function prior and dynamic SFH prior introduced in \citealt{Wang2023}) used to optimize the inference for JWST surveys, can also be properly applied. However, we neglect the uncertainty in the lens model itself, which will be explored in future works.

\subsection{Sample Selection}
We create an initial ``Cosmic Noon'' sample of robustly photometered galaxies using all sources with \texttt{USE\_PHOT=1} \citep[see][]{Weaver2023} and $S/N>10$ in both F150W and F444W between $1<z_{\rm phot}<3$ and $7<\log(M_\star/M_\odot)<11$ using SPS photometric redshifts and stellar masses \citep{Wang2023}, as shown in Table \ref{tab:samplesize}. To ensure accurate $U-V$ colors and recent SFRs, we restrict our sample to sources that have photometric coverage (regardless of $S/N$) in at least one filter blueward of rest-frame 3500 \AA{}. 
Quiescent galaxies are then selected using rotated $UVJ$ coordinates from \cite{Belli2019}:
\begin{align}
    \begin{split}
        &S_Q=0.75~(V-J)+0.66~(U-V)\\
        &C_Q=-0.66~(V-J)+0.75~(U-V).
    \end{split}
\end{align}
Physically, $S_Q$ measures the net slope of a spectrum while $C_Q$ approximates the spectrum's curvature, the difference in slope above and below 4000 \AA{} \citep{Fang2018}. The age of a quiescent galaxy has been shown to increase with $S_Q$ due to the transition from Balmer break- to 4000 \AA{} break-dominated spectra \citep{Whitaker2012,Whitaker2013,Belli2019}. This can be used to infer the median stellar age of a galaxy via:
\begin{align}\label{eqn:t50}
    \log(t_{50}/\rm{yr})=7.03+1.12~S_Q.
\end{align}
In this work, we investigate galaxies with $C_Q>0.49$ \citep[defined as quiescent by][]{Belli2019} and $t_{50}>30$ Myr, indicated by solid black and dotted blue lines, respectively, in the $UVJ$ diagram shown in the left and center panels of Figure \ref{fig:sampleplots}. This extension to the traditional quiescent box has been proven to capture rapidly-quenched and post-starburst galaxies \citep[e.g.,][]{Park2023}.

\begin{figure}
    \centering
    \includegraphics[width=\linewidth]{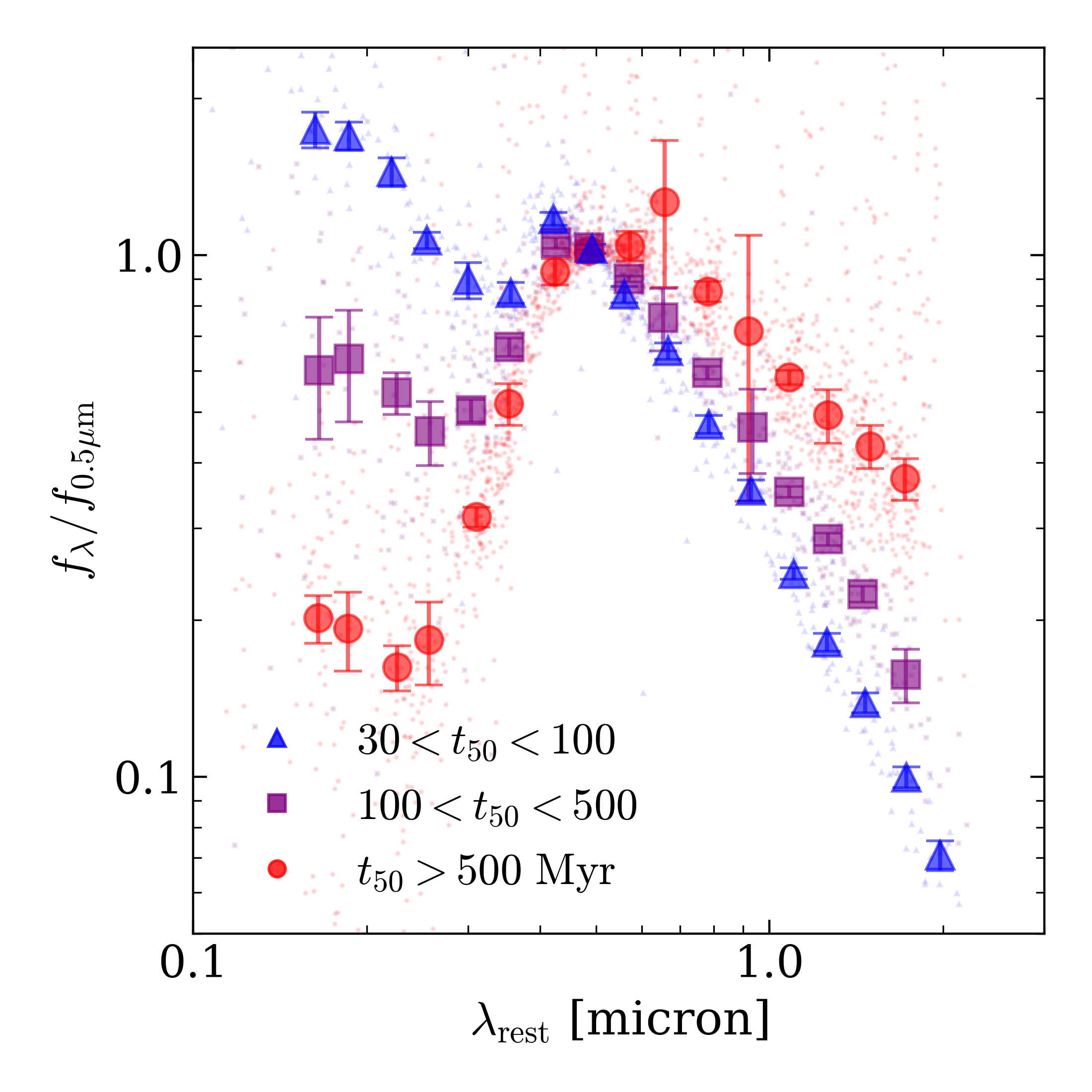}
    \caption{$30<t_{50}<100$ quenched (blue triangles), $100<t_{50}<500$ quiescent (purple squares), and $t_{50}>500$ Myr quiescent galaxies (red circles) have fundamentally different spectral shapes. Composite SEDs are shown with large points while individual photometric data is shown with small points. Error bars show the standard error in the mean. Composite SED points are only shown in wavelength bins that contain $>1\%$ of the total number of photometric points for a given subsample.}
    \label{fig:compositesed}
\end{figure}

In order to remove potential star-forming contaminants from our sample of color-selected quiescent galaxies, we also examine a galaxy's distance from the star-forming main sequence (SFMS) using SPS modeling from \prosb{}. The distance from the SFMS is calculated via 
\begin{align}\label{eqn:dms}
    \deltams{t}(M_\star,z)=\log({\rm SFR_t})-\log({\rm SFR_{MS}}(M_\star,z)),
\end{align}
where t is the lookback time over which the SFH is averaged, and ${\rm SFR_{MS}}(M_\star,z)$ is the SFR of the SFMS at stellar mass $M_\star$ and redshift $z$ from \cite{Leja2022}. Any galaxy more than 0.5 dex below the SFMS at 10 Myr ($\deltams{10}<-0.5$) in our color-selected sample is retained. Selecting quiescent galaxies with $\deltams{100}$ instead of $\deltams{10}$ does not significantly impact our results. We show how $\deltams{10}$ changes with stellar mass for galaxies in our sample in the right panel of Figure \ref{fig:sampleplots}. Galaxies are also colored by their $\deltams{10}$ in the $UVJ$ diagrams shown in the left and center panels of Figure \ref{fig:sampleplots}. 

From the $UVJ$/SFR sample of quiescent galaxies, we compare rest-frame $U-V$ colors measured by \prosb{} and \eazy{} \citep{Brammer2008}. Any outlier sources that differ by $>0.2$ mag in $U-V$ color, which is unphysical if the 4000 \AA{} break is sufficiently sampled, are removed from the sample due to untrustworthy SPS models. 22 total sources of the initial quiescent galaxy sample are removed for outlier $U-V$ colors (6.2\% of the 354 galaxies in row 4 of Table \ref{tab:samplesize}). With color outliers removed, we select 94 galaxies in UNCOVER and 239 galaxies in PRIMER (bold counts in Table \ref{tab:samplesize}). 

\subsection{Separating Quiescent Galaxies by Age}
Lastly, we separate our primary sample into specific types of quiescent galaxies: $t_{50}>500$ Myr quiescent, $100<t_{50}<500$ Myr quiescent, and $30<t_{50}<100$ Myr quenched galaxies, where ``quenched'' is used to indicate galaxies that are observed with colors and measured SFRs consistent with low star-formation activity, but may not be permanently quiescent. $t_{50}>500$ Myr quiescent galaxies are predominantly high-mass ($\log(M_\star/M_\odot)>10$), while the $30<t_{50}<100$ quenched sub-sample is almost exclusively low-mass (Fig. \ref{fig:sampleplots}, right). The required quality selections ($S/N>10$, rest-U coverage) may impact our ability to detect $t_{50}>500$ Myr quiescent galaxies at $\log(M_\star/M_\odot)<9.5$. We note the measured size and structural trends do not change significantly if we apply a \texttt{USE\_PHOT=1} selection only and remove the stringent requirements of $S/N>10$ in F150W and F444W and photometric coverage blueward of 3500\AA{}.  Moreover, the fraction of the sample with  $t_{50}>500$ Myr is roughly unchanged, both in the overall sample and at low masses. Although we expect our source detection strategy to only detect $\sim50\%$ of these these old, red sources at low masses in the first place (discussed more in Section \ref{sec:discuss}), it is clear that our additional quality cuts do not disproportionately remove older quiescent galaxies from the sample.

Final number counts of each sub-population are shown in Table \ref{tab:samplesize}. Figure \ref{fig:compositesed} shows composite spectral energy distributions (SEDs) of each subsample, indicating that each population has distinct spectral shapes. $t_{50}>500$ Myr quiescent galaxies (circles) have strong Balmer breaks with very little ultraviolet (UV) flux and more infrared (IR) flux. $30<t_{50}<100$ Myr quenched galaxies have smaller breaks and more UV flux, indicating their star formation only ceased recently, while $100<t_{50}<500$ Myr quiescent galaxies occupy an intermediate region between the two.

\section{Analysis}\label{sec:methods}
We adapt the GALFIT \citep{Peng2002,Peng2010} one-component S\'ersic fitting methods from \cite{vanderWel2012}. Galaxies are fit in both the F150W and F444W filters: F150W is chosen due to its high spatial resolution and comparable wavelength coverage to HST/WFC3 F160W, while F444W provides the longest wavelength coverage available. All science images are processed in the same way as follows.  

Cutouts of each source are taken from the science, weight, exposure time, and segmentation images. The cutouts have sides of length $7\times R_{\rm{Kron,circ}}$ or 150 pixels, whichever is larger. Sources where $>40\%$ of the cutout or the central pixel is empty (zero weight) are removed. Of the 333 galaxies in the sample, 2 (0.6\%) are removed by this cut, both of which are located on the edge of the mosaic. The error at a given pixel, $\sigma_i$, is estimated using a combination of sky background variance ($1/w$) and Poisson noise ($f/t_{\rm exp})$:
\begin{align}
    \sigma_i=\sqrt{\frac{1}{w_i}+\frac{f_i}{t_{\rm{exp},i}}},
\end{align}
where $w_i$, $f_i$, and $t_{\rm{exp},i}$ are the weight, flux, and exposure time of a pixel. To mask all other sources in the cutout, nearby-object masks are created from the segmentation map. Empirical JWST point-spread functions (PSFs) are built using stacks of unsaturated stars selected directly from the mosaic, as in \cite{Weaver2023}, and normalized to reported encircled energies from calibration resources\footnote{\url{https://jwst-docs.stsci.edu/jwst-near-infrared-camera/nircam-performance/nircam-point-spread-functions}} at 4\arcsec{}.

\begin{figure*}[ht!]
    \centering
    \subfloat{\includegraphics[width=0.205\linewidth]{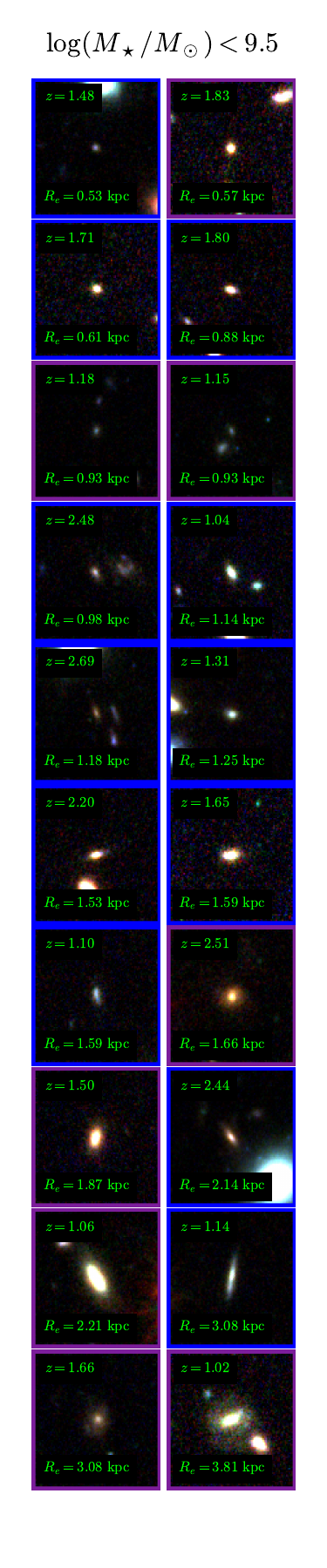}}
    \subfloat{\includegraphics[width=0.5\linewidth]{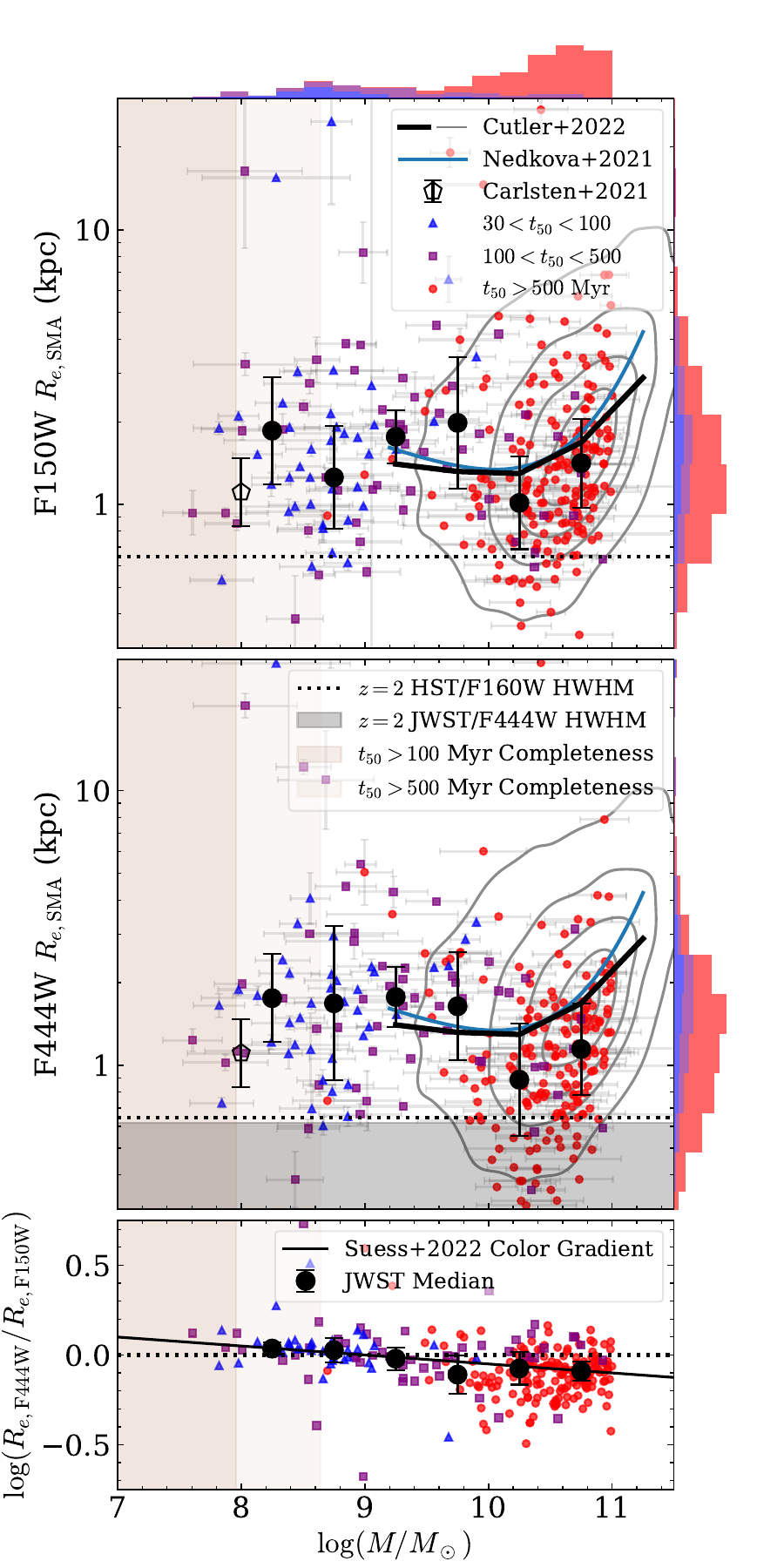}}
    \subfloat{\includegraphics[width=0.205\linewidth]{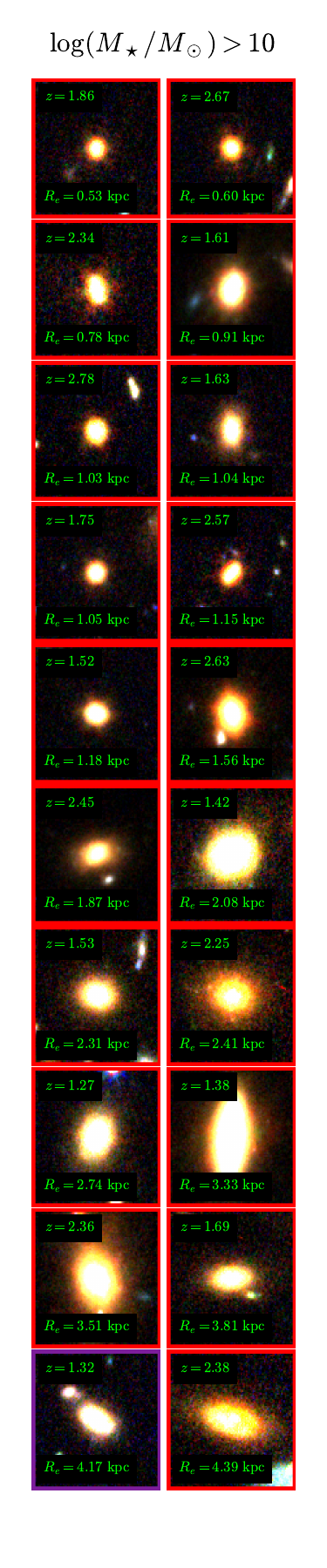}}
    \caption{
    Center top and middle: The quiescent galaxy size-mass relation changes significantly at low masses. Quiescent size-mass relations in both JWST/NIRCam F150W (top) and F444W (middle) are shown with blue triangles, purple squares, and red circles indicating the individual sizes of $30<t_{50}<100$, $100<t_{50}<500$, and $t_{50}>500$ Myr quiescent galaxies, respectively. Light gray error bars indicate the 1$\sigma$ uncertainty on size and mass for galaxies in the sample. Solid gray contours and thick black lines indicate the $1<z<3$ HST/WFC3 F160W sizes of galaxies from COSMOS-DASH \citep{Cutler2022}, while cyan lines represent the $1.5<z<2.0$ quiescent size-mass relation from \cite{Nedkova2021}. Open pentagons show the median size-mass of local dwarf satellites from \cite{Carlsten2021}. Thin black dotted lines show the F160W HWHM and black shading shows the F444W HWHM at $z=2$. The 50\% mass-completeness detection limits of the combined PRIMER and UNCOVER sample for $t_{50}>100$ and $>500$ Myr quiescent galaxies is indicated with dark and light brown shading, respectively. Marginal histograms show the 1D distributions of galaxy mass and size.
    Center bottom: The ratio of F444W to F150W sizes and exhibits comparable trends to the \cite{Suess2022} color gradients (thin black line).  
    Left and Right: F444W-F277W-F150W color images for example low-mass (left) and massive (right) galaxies in our sample highlight the morphological differences between these mass regimes. Galaxies are ordered from top-left to bottom-right by increasing F150W size (in kpc) and the boundaries of the cutouts are highlighted in blue, purple, or red according to the median stellar age of the galaxy (as in Figure \ref{fig:compositesed}).
    }
    \label{fig:sizemasstot}
\end{figure*}

The science, error, and nearby-object mask cutouts, as well as the empirical PSF, are provided to \galfit{} and used to determine the best-fit S\'ersic model with free parameters including the object centroid ($x_0$, $y_0$), total magnitude ($m$), semi-major axis effective (half-light) radius ($R_e$), S\'ersic index ($n$), axis ratio ($q$), and position angle ($\theta$). No additional sky background pedestal is fit as we find JWST performs better with external background subtractions \citep[as opposed to HST,][]{Haussler2007,Cutler2022}, with 5 additional sources being successfully fit when background subtracting is disabled. Standard constraints are imposed on the magnitude ($\pm3$ mag from the photometric catalog value), radius ($0.01<R_e<400$ pixels), S\'ersic index ($0.2<n<10$), and axis ratio ($0.0001<q<1$). A bootstrapping method is used to compute uncertainties on the modeled parameters. We create 100 realizations of each galaxy by applying a random perturbations drawn from the background sky variance to the science image. Each realization is fit with \galfit{} and the $1\sigma$ uncertainty of each parameter is computed via the median absolute deviation in the resulting distributions.

Sources that have poor mosaic coverage (see above), have best-fit parameters at the constraint values, or cannot be fit by \galfit{} \citep[i.e. \galfit{} \texttt{FLAG}$\geq2$ in][]{Cutler2022} are removed from the sample. In total, 41 galaxies (12\% of the 333 galaxy sample) are removed from the sample due to inadequate coverage or \galfit{} fitting (3 at $30<t_{50}<100$, 13 at $100<t_{50}<500$, and 25 at $t_{50}>500$ Myr, with most $t_{50}>500$ Myr quiescent galaxies at $\log(M_\star/M_\odot)>10$). We also remove 5 sources (1.7\% of the 292 remaining galaxies) with large uncertainty in stellar mass ($\delta\log(M_\star/M_\odot)>1$ dex). To account for the effects of lensing, UNCOVER sizes are scaled by $1/\sqrt{\mu}$ \citep[where $\mu$ is the magnification as reported in][]{Furtak2023}. Sources with high magnification $\mu>3$ are removed from the sample due to possible impacts on the galaxy's size and shape. This removes 7 of the 292 galaxies left after removing bad \galfit{} fits (2.4\%), of which one is a low-mass, $t_{50}>500$ Myr sources. If we remove the UNCOVER sample entirely, the overall trends are statistically unchanged, suggesting that lensing effects are minimal. This also suggests that the additional depth of UNCOVER (29.21 ABmag compared to 28.17 ABmag in PRIMER) is unnecessary for smooth (e.g. S\'ersic) light profile-fitting. We find observed sizes with shallower data increases scatter overall, but does not introduce biases in the median relation. However, more in depth modeling is likely necessary to unambiguously demonstrate this effect.

\begin{figure*}[ht!]
    \centering
    \subfloat{\includegraphics[width=0.5\linewidth]{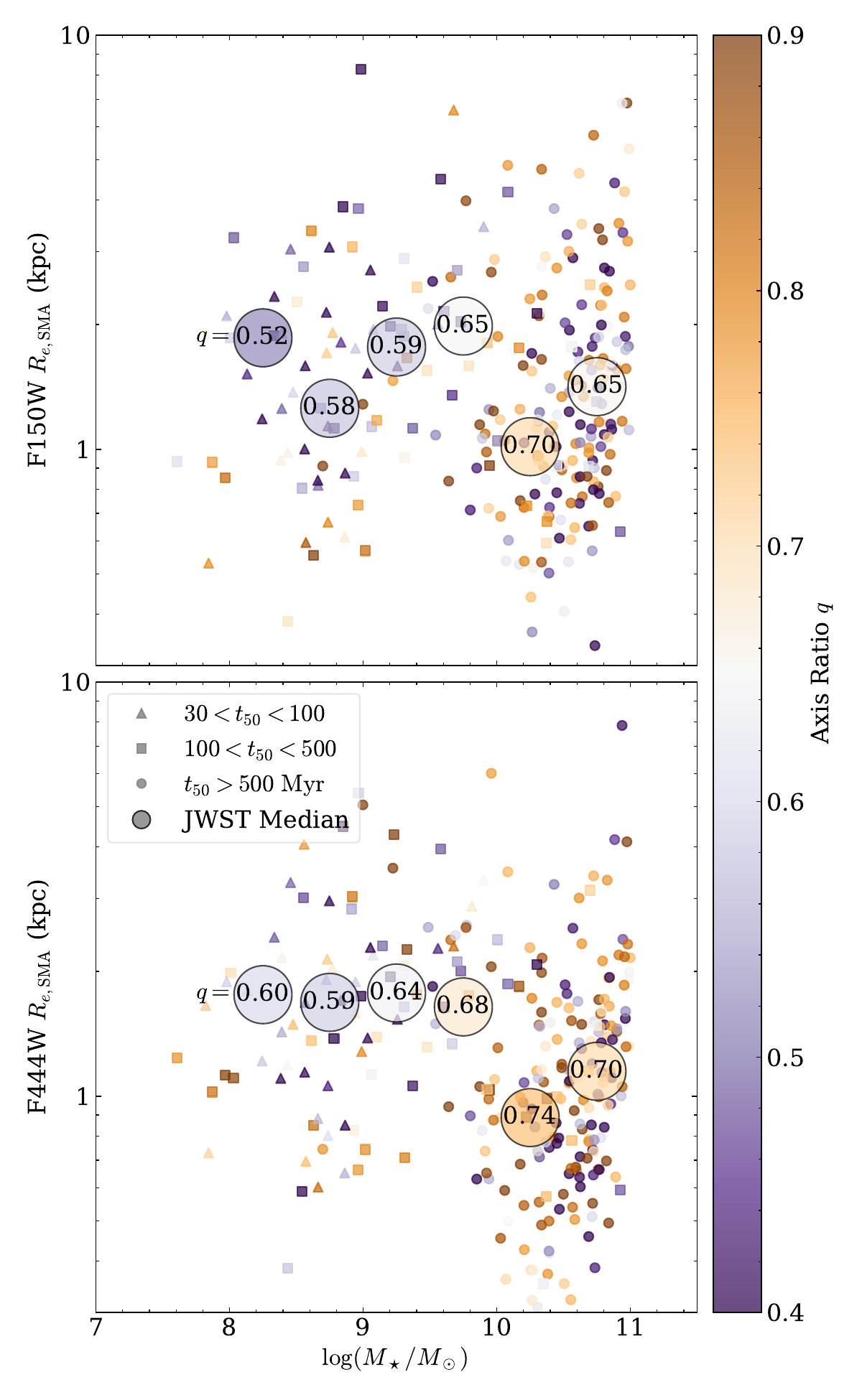}}
    \subfloat{\includegraphics[width=0.5\linewidth]{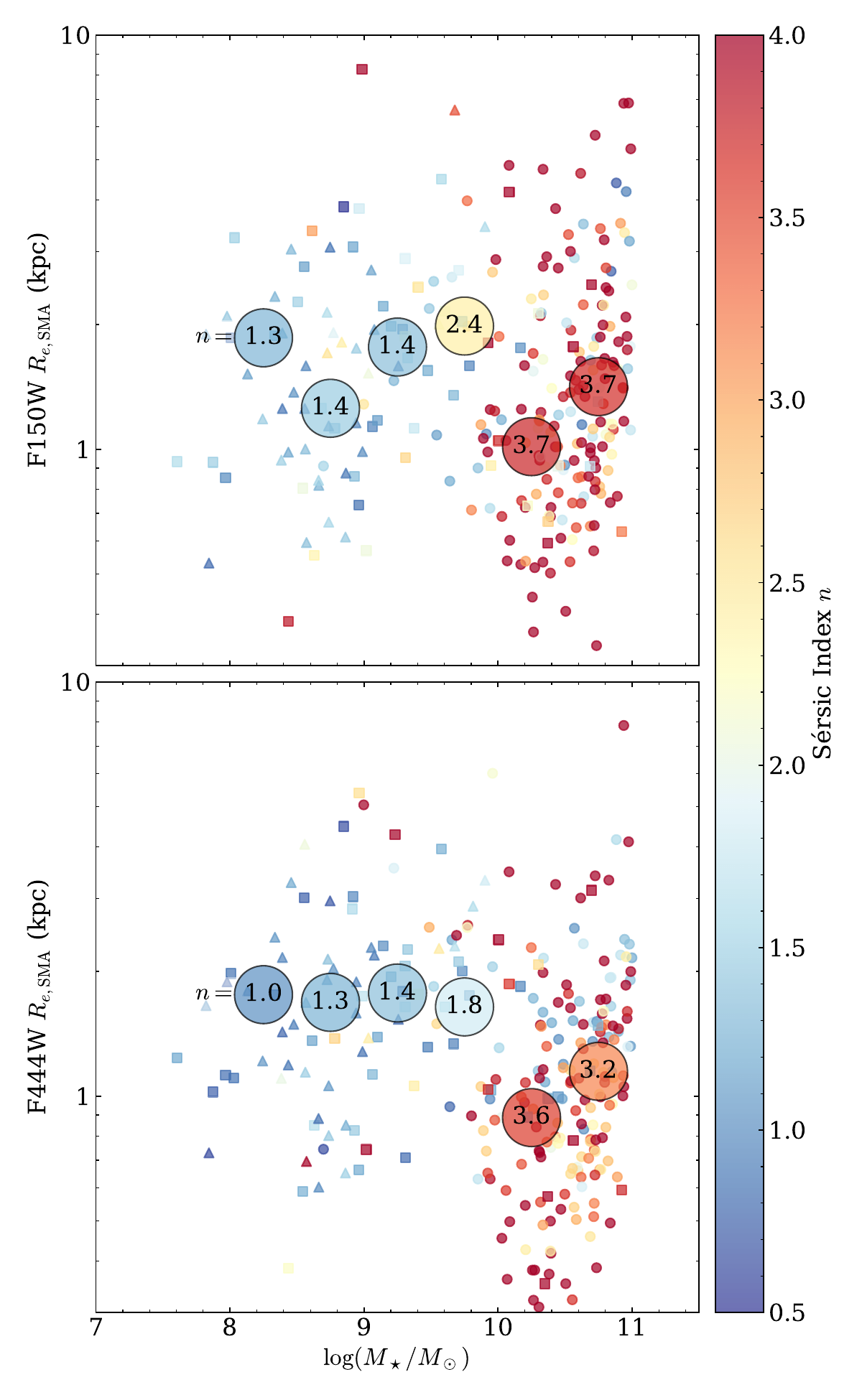}}
    \caption{The size-mass relation for galaxies in our sample colored based on the axis ratio (left) and S\'ersic index (right) for both F150W (top) and F444W (bottom). Galaxies in the flattened part of the size-mass relation have smaller axis ratios and S\'ersic indices. Individual points have the same shape symbols as Figure \ref{fig:sizemasstot}. Large circles indicate the median size-mass for a given stellar mass bin, color-coded by the median axis ratio or S\'ersic index. The median axis ratio and S\'ersic index are also shown numerically in each point. Typical error bars on axis ratio are $\pm0.10$ at $\log(M_\star/M_\odot)<10$ and $\pm0.04$ at higher masses. For the S\'ersic index, error bars are typically $\pm0.4$.}
    \label{fig:sizemassqn}
\end{figure*}

\section{Results}\label{sec:results}
\subsection{The Quiescent Galaxy Size-Mass Relation}\label{sec:sizemass}
If the quiescent size-mass relation really does flatten at low masses, we should expect to see it clearly in both F150W and F444W. In Figure \ref{fig:sizemasstot}, we show the quiescent size mass relation in both F150W (top) and F444W (middle), with a comparison of the sizes of both filters (bottom). There is indeed a significant flattening of the size-mass relation at $\log(M_\star/M_\odot)<10$ that is present in both filters. The median size in the flattened region of the size-mass plot of 1.6 kpc is roughly the same in both filters \citep[as in, e.g.,][]{Nedkova2021,Cutler2022}, as shown in the bottom panel of Figure \ref{fig:sizemasstot}, falling to a minimum at $\log(M_\star/M_\odot)\sim10.3$. The trend then increases with a stronger mass-dependence at higher masses \citep[as in, e.g.,][]{vanderWel2014,Mowla2019,Cutler2022,Ito2023}. These trends are apparent well above the JWST resolution limit and the mass completeness limits for different age quiescent galaxies (black and brown shading, respectively).

Also apparent is a bimodality in the overall quiescent galaxy stellar mass distribution of this sample (marginal histogram in Figure \ref{fig:sizemasstot}). This behaviour, combined with the disjoint size-mass relations, indicates low- and high-mass quiescent galaxies may exist in two separate distributions altogether, more in line with proposed star-forming and quiescent growth tracks \citep[e.g., Figure 28 in][]{vanDokkum2015}. We discuss the possible physical causes for the dip in size at $\log(M_\star/M_\odot)\sim10.3$ as well as the distinct distributions of size and mass in Section \ref{sec:discuss}. 

There exists a slight difference between earlier HST median sizes (e.g., COSMOS-DASH, \citealt{Cutler2022}; CANDELS/HFF,  \citealt{Nedkova2021}) and this JWST sample: the median high-mass ($\log(M_\star/M_\odot)>10$) size is smaller in JWST, especially in F444W. With deeper JWST data we can test if this tension is the result of biases in the shallower, noisier HST data. To test this, we remeasure the sizes of the galaxies in our sample with increased noise comparable to the depth of COSMOS-DASH observations ($m_{5\sigma}\sim25$). We find that while the scatter noticeably increases and fewer low-mass galaxies are successfully fit, there are no biases in the median size that would explain the differences with HST. This effect is also not likely due to the increased spatial resolution of JWST relative to HST. There are noticeably fewer high-mass quiescent galaxies above the HST size-mass relations, so removing sources below the HST HWHM won't significantly increase the median JWST sizes. Moreover, this effect is most significant in F444W, which has a comparable PSF HWHM to F160W. 

Instead, we find that the large discrepancy in F444W is likely due to the negative color gradients in massive quiescent galaxies (solid black line in Figure \ref{fig:sizemasstot}, center-bottom), where sources are more compact in the rest-frame near-infrared than the rest-UV/optical \citep{Suess2022,vanderWel2023}. These color gradients are likely due to physical differences in these galaxies when observed at rest-UV and rest-optical wavelengths, as opposed to resolution effects: on average F150W sizes are larger than F444W sizes by 0.08 dex when F150W imaging is PSF-matched to the broader spatial scales of F444W, which is comparable to the mean color gradient at native F150W resolution (0.07 dex). The small offset in F150W may be the result of the additional SFR selection we impose on our quiescent galaxy sample: we can recover the high-mass relations of \cite{Cutler2022} and \cite{Nedkova2021} by not implementing SFR cuts, suggesting earlier works possibly had contamination from larger star-forming galaxies. 

\vspace{0.5cm}
\subsection{Trends with Structural Parameters}
Figure \ref{fig:sizemassqn} shows the JWST size-mass relations from Figure \ref{fig:sizemasstot} color-coded by axis ratio, $q$ (left), and S\'ersic index, $n$ (right). In both filters, we see clear trends with $n$. In the low-mass, flattened region of the size-mass relation, galaxies have much smaller S\'ersic indices ($n<2$). At the high-mass, steeply growing end, S\'ersic indices are in the traditionally ``elliptical'' regime ($n>2.5$). At $\log(M_\star/M_\odot)>9.5$, axis ratios are generally higher than at lower masses. This higher median $q$ in this older, more massive population, combined with the larger S\'ersic indices is likely indicative of a spheroidal population, compared to the lower-mass sample that appears consistent with a more disk-like population. This dichotomy is also apparent in the RGB images of individual galaxies (Figure \ref{sec:sizemass}, left and right). Unusually, in all mass bins, F444W S\'ersic fits find higher values for $q$ and lower values for $n$ than F150W, contrary to \cite{Martorano2023}. However, \cite{Martorano2023} find very weak trends between wavelength and structural parameters, and their sample is limited to a relatively small number of massive galaxies at $z>1.5$. Moreover, the scatter in the median $q$ and $n$ prevent us from making strong claims about the dependence of these structural parameters on observed wavelength.

\begin{figure*}[ht!]
    \centering
    \includegraphics[width=1\linewidth]{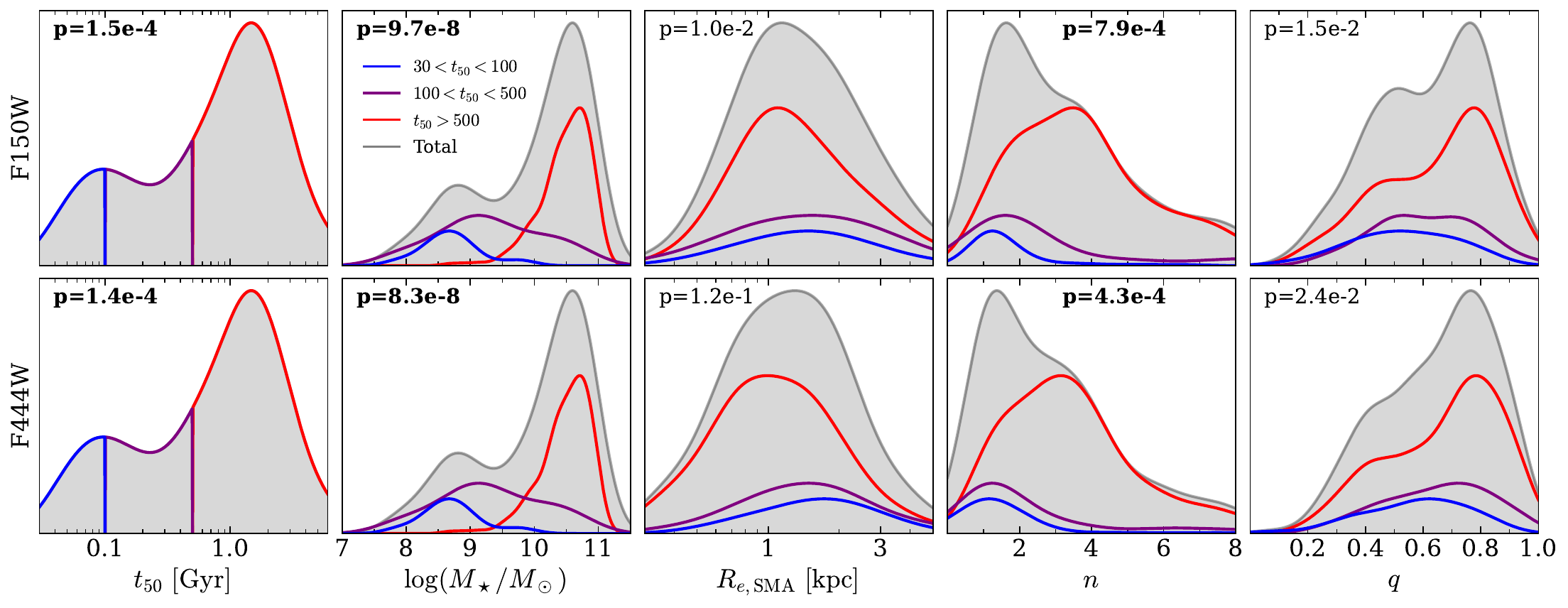}
    \caption{In general, physical parameters differ dramatically between low- and high-mass quiescent galaxies, though $R_e$ and $q$ alone are not sufficient to separate these samples. Smoothed histograms showing distributions of age, mass and structural parameters ($R_e$, $n$, $q$) for F150W (top) and F444W (bottom). Individual distributions of $30<t_{50}<100$, $100<t_{50}<500$, and $t_{50}>500$ Myr quiescent galaxies are shown with blue, purple, and red curves, respectively, while the total distribution is shown with a grey shaded curve. P-values for total-sample KS tests are shown, with bold values indicating that the parameter distribution is significantly different from a single Gaussian distribution.}
    \label{fig:corners}
\end{figure*}

Figure \ref{fig:corners} compares the $t_{50}$, stellar mass, $R_e$, $n$, and $q$ distributions for $30<t_{50}<100$, $100<t_{50}<500$ and $t_{50}>500$ Myr quenched/quiescent galaxies. Quiescent galaxies with low S\'ersic indices ($n<2.5$) have been observed with a roughly flat distribution of axis ratios \citep[e.g.,][]{Chang2013,Cutler2022}. However, we see a peak in the axis ratios of $30<t_{50}<100$ quenched and $100<t_{50}<500$ Myr quiescent galaxies (which predominately have $n<2.5$) at $q\sim0.6$ and $0.7$, respectively. These distributions, combined with the smaller median S\'ersic indices we measure, likely indicate that the younger ($t_{50}<500$ Myr), low-mass quiescent galaxies are oblate and disky. This is in agreement with \cite{Tan2022}, who find that disk-like structures dominate the low-mass ($8.5<\log(M_\star/M_\odot)<9.5$) quiescent galaxy population in the Hubble Frontier Fields. For $t_{50}>500$ Myr quiescent galaxies, the $q$ distribution peaks significantly at $q\sim0.8$. This peak also coincides with a higher S\'ersic index ($n>3$), which strongly suggests these galaxies are predominantly spheroidal.

\section{Discussion}\label{sec:discuss}
The distinct size-mass relations observed in quiescent high- and low-mass galaxies (no size evolution at low masses and increasing size with mass high masses) potentially indicates two separate evolutionary paths. These paths may be caused by separate quenching mechanisms and/or different physical processes driving size growth.

\subsection{Structural Evolution at High Masses}
At high masses, compaction is often associated with quenching \citep[e.g.,][]{Cheung2012,Barro2017,Whitaker2017,Lee2018,Ji2022}; several studies have even proposed morphological changes, specifically bulge formation and compaction, as potential mechanisms for quenching in massive elliptical galaxies \citep[e.g.,][]{Martig2009,Tacchella2018,Tadaki2020}. These structural changes would result in the smaller $R_e$ and larger $n$ seen in $10<\log(M_\star/M_\odot)<11$ quiescent galaxies in Figures \ref{fig:sizemasstot} and \ref{fig:sizemassqn}.

However, the steeper size-mass slope of massive quiescent galaxies \citep[e.g.,][]{Mowla2019} cannot be explained by compaction/bulge formation alone. Moreover, it isn't clear why the disjoint transition between these populations occurs at $\log(M_\star/M_\odot)\sim10.3$. These effects could be explained by a change in the dominant mechanism behind size growth at this transition mass. At lower masses, galaxies can only grow in size via star formation (prior to quenching), whereas at higher masses, dry mergers rapidly increase the size of previously-quenched galaxies \citep{vanDokkum2015}. 

At $1<z<3$, the quiescent mass function peaks at $\log(M_\star/M_\odot)\sim10.5$ \citep{Muzzin2013,Santini2022,Weaver2023a}, which is roughly the mass of the size-growth transition found herein. Since quiescent galaxies at this mass are most abundant, they are the most likely candidates for mergers and a clear starting point for the steeper growth found at the high-mass end. Galaxies below this mass are also more likely to merge with a higher mass galaxy, which prevents steep size growth from occurring at $\log(M_\star/M_\odot)<10.5$ as they are subsumed. Merger-dominated growth at high masses could also explain the dip in the quiescent size-mass relation at $\log(M_\star/M_\odot)\sim10.3$: galaxies in this mass regime may be in a compact, non-virialized state following the merger. 

In total, we speculate that two factors explain the shape of the quiescent size-mass relation at high masses: compaction associated with mass-quenching and merger-driven growth of massive galaxies. Compaction is responsible for the clear minimum in quiescent galaxy size at $\log(M_\star/M_\odot)\sim10.3$, while mergers explain the steady increase in size at larger masses. The shape of the quiescent galaxy mass function is also crucial, as it prevents merger-driven growth at lower-masses (since $\log(M_\star/M_\odot)\sim10.5$ quiescent galaxies are most abundant).

\subsection{Structural Evolution at Low Masses}
At low masses, quenching could instead be primarily driven by mechanisms that mostly leave structure and size intact, similar to star-forming galaxies at comparable masses. Environmental quenching such as stripping of gas in and around a low-mass galaxy \citep[e.g.,][]{Weinmann2006}, is a viable explanation. \cite{Yoon2023} find at nearby redshifts ($0.01<z<0.04$), the quiescent size-mass relation flattens at $\log(M_\star/M_\odot)<10$ for galaxies in denser environments, while isolated low-mass quiescent galaxies are smaller and their size-mass relation does not flatten. Low-mass quiescent galaxies in high-density environments also have smaller S\'ersic index and more disky structure \citep[e.g.,][]{Carlsten2021,Yoon2023}, which could be due to environmental gas stripping without significant morphological changes. Moreover, recent studies with JWST have found that low-mass ($\log(M_\star/M_\odot)<9.5$) post-starburst galaxies at $z>3$ are preferentially found in overdense environments \citep{Alberts2023}. The existence of known cosmic noon overdensities in the observed fields (COSMOS: \citealt{Spitler2012,Chiang2014}, UDS: \citealt{Chuter2011,Guaita2020}) suggests that environmental quenching is a potential explanation for the resulting structure of low-mass quiescent galaxies in our sample. 

As a test for this mechanism, we compare the redshifts and positions of the 58 $\log(M_\star/M_\odot)<10$ quiescent galaxies in PRIMER-COSMOS to 39 reported cosmic noon overdensities in the COSMOS field from \cite{Spitler2012} and \cite{Chiang2014}. 17 of our sources were found at similar redshifts ($\Delta z<0.1$) to these overdensities, of which 2 were within 5\arcmin{} of the overdensity centers. While the compiled sample of COSMOS overdensities is not exhaustive, there is little association between our low-mass quiescent galaxies and these sources. However, we would likely need confirmed spectroscopic redshifts to robustly test for associations, as the photometric redshifts in our total sample have a typical uncertainty of $\delta z=0.17$, almost twice as large as the scatter in our test.

While simulations find that low-mass satellites of the Milky Way and M31 can be efficiently quenched \citep{Fillingham2018}, these studies also suggest that low-mass field galaxies are unlikely to be quenched by galaxy-galaxy interactions, and that they are more likely quenched by efficient feedback mechanisms \citep[e.g.,][]{Fitts2017}. The choice of feedback model has also been shown to impact the overall size-mass relation at low masses \citep[e.g.,][]{Pillepich2018}, with the caveat that the sample isn't separated into star-forming and quiescent populations. At higher redshifts, supernovae have been proposed as potential driver of low-mass galaxy quenching, though they have been shown to not impart sufficient energy to halt star formation in the \cite{Looser2023} or \cite{Strait2023} high-redshift mini-quenched galaxy candidates \citep{Gelli2023b,Gelli2023}. Quenching via environmental interactions would result in minimal structural changes to the stellar light of low-mass galaxies, while stellar feedback quenching makes galaxies larger \citep{Ubler2014}. This lack of significant structural changes in an environmental or feedback quenching scenario would explain why low-mass quiescent galaxy sizes are similar to low-mass star-forming galaxies, as well as why their S\'ersic indices are closer to an exponential disk light profile. Our data is consistent with low-mass quenching being driven by stellar feedback and environmental interactions, and while we can only speculate, it seems plausible that both processes may play a role in low-mass quenching.

The stark flatness of the low-mass quiescent size-mass relation is potentially due to a combination of halted galaxy growth after quenching (in the absence of mergers) and progenitor bias \citep[which is pronounced for low-mass galaxies, e.g.,][]{Ji2023}, as suggested by the age gradient in the low-mass quiescent galaxy population. In the flattened region, more massive galaxies are older and thus formed earlier in the universe, which makes them smaller and more dense than more recently formed galaxies \citep{Mo2010,Ji2023}. Meanwhile, the younger, low-mass galaxies formed later and are intrinsically larger than older galaxies when they quench. In the absence of mergers at $\log(M_\star/M_\odot)<10$, the sizes of quiescent galaxies should remain the same, meaning the young, low-mass galaxies have roughly the same size as the older, more massive galaxies.

\subsection{Two Populations of Quiescent Galaxies}
The potential existence of different evolutionary paths for high- and low-mass quiescent galaxies is highlighted further by a significant paucity in the number of $1<z<3$ quiescent galaxies between $9\lesssim\log(M_\star/M_\odot)\lesssim10$ (mass histograms in Figures \ref{fig:sizemasstot} and \ref{fig:corners}). This deficit is likely a real physical effect and not an artifact of our stellar population modeling or sample selection, as it exists separately in UNCOVER and PRIMER (for larger, less curated quiescent galaxy selections) and using stellar population parameters derived independently with \eazy{} and \pros{}. Previous studies hint at a two peaked distribution in stellar-mass both locally \citep{YPeng2010} and at increasingly higher redshifts \citep{Ilbert2013,Davidzon2017,Santini2022,Weaver2023a}. In particular, the COSMOS2020 quiescent mass function from \cite{Weaver2023a} finds a local minimum in the number of quiescent galaxies between $9\lesssim\log(M_\star/M_\odot)\lesssim10$, however is only complete to $\log(M_\star/M_\odot)\sim9$ at $z\sim1.5$. Now with the deep, space-based data from JWST we are able to observe this behavior out to $z\sim3$ (see Pan et al. in prep).

To confirm the existence of distinct size-mass relations in these regimes, we run Spearman correlation tests for $t_{50}>500$ Myr and $t_{50}<500$ Myr galaxies separately, as these selections are roughly equivalent to the two mass ranges considered. For sizes measured in both filters, the $t_{50}>500$ Myr sample has a clear positive correlation ($r\sim0.2$, $p\ll0.01$), while the $t_{50}<500$ Myr sample finds no correlation, confirming the flat relation we see at low-masses. If we instead test for correlation in subsamples selected by stellar mass ($\log(M_\star/M_\odot)<10$ and $\log(M_\star/M_\odot)>10$), we find stronger correlations at high masses ($r\sim0.3$), while the low-mass regime still has no correlations. As a baseline, we also perform Spearman tests for the entire sample of quiescent galaxies. There is no correlation in the overall sample in F150W, however in F444W we find a slight negative correlation ($r=-0.16$, $p\sim0.01$). This is likely driven by smaller sizes at the massive end as a result of the negative color gradient observed at high masses.

\begin{figure}
    \centering
    \includegraphics[width=\linewidth]{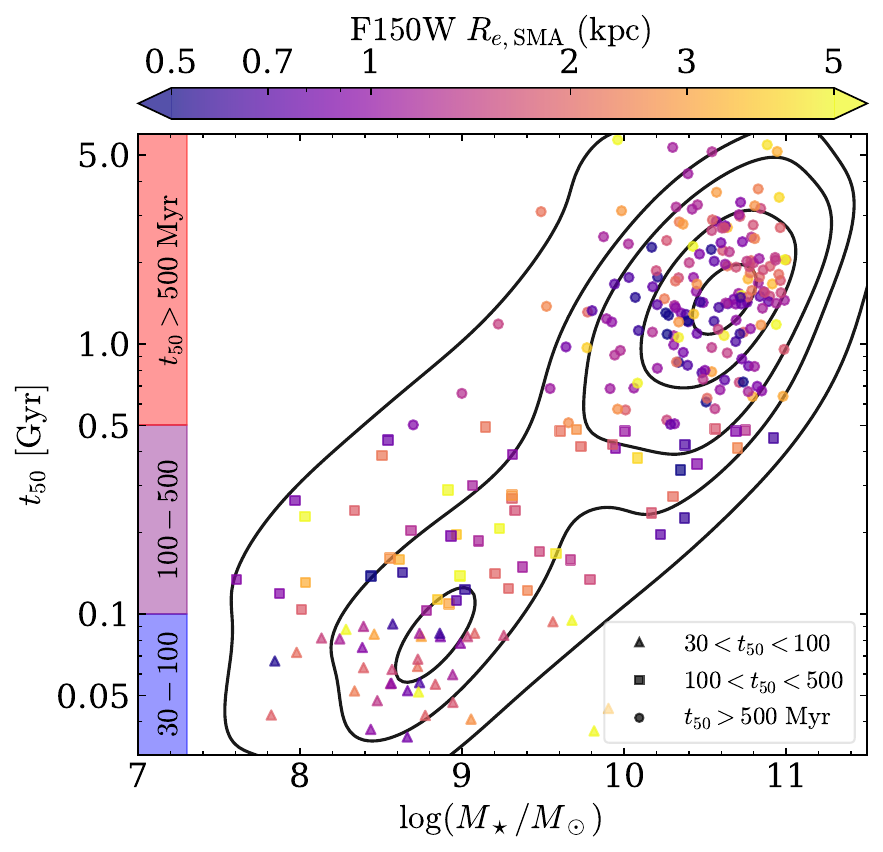}
    \caption{Quiescent galaxies at cosmic noon separate into two distinct populations in age-mass space. Median stellar age is computed from the color-estimated $t_{50}$ (Eqn. \ref{eqn:t50}). Triangles, squares and circles (along with blue, purple, and red bars on the left) indicate the $30<t_{50}<100$ quenched, $100<t_{50}<500$ quiescent, and $t_{50}>500$ Myr quiescent galaxy subsamples, respectively. Points are colored according to their F150W effective radii.}
    \label{fig:mass-age}
\end{figure}

The separation of low- and high-mass quiescent galaxies into two populations becomes clearer in age-mass space. Figure \ref{fig:mass-age} shows the sample of quiescent galaxies in age-mass space (where median age is estimated from Equation \ref{eqn:t50}). Both the density contours and individual points suggest there are two populations: young, low-mass ($\log(M_\star/M_\odot)<9.5$) quiescent galaxies and older, massive ($\log(M_\star/M_\odot)>10$) quiescent galaxies. The young population is predominantly composed of the $30<t_{50}<100$ and $100<t_{50}<500$ Myr subsamples, with F150W effective radii (indicated by the color of the points) scattered roughly around $\sim$ 1 kpc. Conversely, the massive population has a significant number of galaxies with $R_{e,{\rm F150W}}<1$ kpc, though in general the populations are not effectively separated by size alone.

To robustly confirm the existence of two separate populations of quiescent galaxies, we compute the p-values of Kolmogorov-Smirnov (KS) Tests for a number of physical and structural parameters for the overall sample of quiescent galaxies, as shown in Figure \ref{fig:corners}. Structural parameters whose distributions deviate significantly from a single Gaussian distribution will have p-values $\ll0.01$. For age, mass, and S\'ersic index, 1D KS tests indicate these parameters are bimodally distributed. However, we are unable to conclude size and axis ratio deviate from a single Gaussian distribution for the entire sample of quiescent galaxies. Lastly, a 2D KS Test \citep{Peacock1983} finds that the age-mass distribution is inconsistent with a single Gaussian distribution, indicated by a p-value significantly less that 0.01 ($10^{-11}$), as is apparent in Figure \ref{fig:mass-age}.

The age-mass separation of these two populations could be a sign that low-mass quenching mechanisms, such as environmental interactions or feedback, become less effective at $9<\log(M_\star/M_\odot)<10$, making galaxies at $\log(M_\star/M_\odot)<9$ more abundant. At slightly higher masses ($\log(M_\star/M_\odot)>10$), merger-driven growth and mass quenching ensures that these galaxies are more abundant. Moreover, the mass where these two populations diverge coincides with several other significant transitions in galaxy evolution. At $\log(M_\star/M_\odot)\sim10.3$, observations show the SFMS slope flattens \citep[e.g.,][]{Whitaker2012,Leja2022}, star-forming galaxies transition from mostly unobscured to dusty \citep{Martis2016}, and the quiescent mass function peaks \citep[e.g.,][]{Muzzin2013,Santini2022,Weaver2023}, and simulations indicate the stellar-halo mass relation and star-formation efficiency peak \citep[e.g.,][]{Behroozi2013,Genel2014}. The fact that several well-studied relations in galaxy evolution, including the quiescent size-mass relations presented herein, see key changes occur at $\log(M_\star/M_\odot)\sim10.3$ is strong evidence that galaxy formation differs dramatically above and below this characteristic mass, likely due to different physics playing a role in each regime.

\subsection{Caveats}
One feature of our results is a lack of $t_{50}>500$ Myr quiescent galaxies at low masses ($\log(M_\star/M_\odot)<9.5$). This is potentially a consequence of our ``flat'' selection function: older galaxies that are bright in F444W only are likely to be missed. We estimate the mass completeness of $t_{50}>100$ and $>500$ Myr quiescent galaxies (brown shading in Fig. \ref{fig:sizemasstot}) by scaling the mass and detection (F277W+F356W+F444W) flux of these galaxies to low masses (assuming a constant mass-to-light ratio). At $\log(M_\star/M_\odot)\sim8.6$, we find our sample requirement of $S/N>3$ only detects 50\% of all $t_{50}>500$ Myr quiescent galaxies. Taking instead the result at face value, one explanation is that low-mass, $t_{50}<500$ Myr quenched galaxies are the progenitors of today's dwarf ellipticals, which are roughly 8-12 Gyr old on average \citep[e.g.,][]{Rakos2001,Jerjen2004}, comparable to the range of lookback times from $1<z<3$. If this population began to form at these redshifts, it may explain why we do not yet see older, low-mass quiescent galaxies: the progenitor populations don't exist at higher redshifts. Similarly, the mechanisms driving low-mass galaxies to quench and become more massive may be rapidly moving quenched galaxies out of the low-mass, disky population and into the massive, spheroidal population on timescales $<500$ Myr. For example, if environmental quenching dominates at low masses, perhaps those interactions are also coupled with more frequent mergers, which would rapidly increase the mass of galaxies and move them out of the low-mass population. Alternatively, the potential existence of so-called ``mini-quenched galaxies'' may explain why $t_{50}>500$ Myr, low-mass quiescent galaxies are uncommon in our sample: they may be experiencing a temporary quenching event and thus rejuvenate before they reach advanced ages. 

While the $30<t_{50}<100$ Myr quenched population in our sample appears to be quenched on at least 10 Myr timescales, we cannot rule out that these galaxies may be lower-redshift counterparts of the mini-quenched galaxies recently identified at high redshifts \citep{Looser2023b,Strait2023}, in which case they will likely resume star formation in the near future. Existing theoretical studies of $z>4$, $\log(M_\star/M_\odot)<9$ galaxies predict that very bursty SFHs are needed to explain observations \citep[e.g.,][]{Angles-Alcazar2017,Faucher-Giguere2018,Ma2018,Sun2023,Dome2023}, but similar studies at lower redshifts do not exist. Moreover, stellar population modeling from broadband photometry alone is largely insufficient to resolve burst durations on time scales shorter than 100 Myr \citep[e.g.,][]{Suess2022a,Wang2023c}, whereas one would need <40 Myr time scales in order to separate candidate mini-quenched galaxies from recently quenched galaxies \citep[e.g.,][]{Dome2023}. Spectroscopy of a sample of these recently quenched galaxies would offer more accurate SFH reconstruction on shorter timescales and allow us to identify observables that isolate mini-quenched galaxies from other low-mass galaxies.

\section{Summary}

The distinct size-mass relations, structural measurements, and mass-age distributions for low- and high-mass quiescent galaxies support the idea that low-mass quiescent galaxies differ dramatically from their higher mass counterparts, regardless of the specific physical processes involved. Below $\log(M_\star/M_\odot)\sim10.3$, quiescent galaxies have a median $R_e\sim1.6$ kpc, irrespective of their stellar mass which unambiguously transitions to the steeper slope of the massive quiescent size-mass relation seen in previous studies \citep[e.g.,][]{Mowla2019,Nedkova2021,Cutler2022}. This shift in mass dependence is also accompanied by a dramatic change in the median S\'ersic index and axis ratio distribution, indicating that this change is associated with galaxies shifting from disky to more elliptical morphologies. The separation of these populations around $\log(M_\star/M_\odot)\sim10.3$ may indicate that these distinct classes of quiescent galaxies are related to the overall dichotomy between high- and low-mass galaxy formation, as suggested by studies of mass functions, global star-formation rates and efficiencies, and halo-to-stellar mass ratios. In the future, investigating environments around low-mass quiescent galaxies will be necessary for robustly tying their quenching to environmental interactions, while medium band and spectroscopic surveys will be crucial to directly probe feedback signatures. Moreover, spectroscopic follow up of potential mini-quenched candidates (low-mass, recently quenched galaxies with rapid changes in SFR), as well as theoretical predictions for this population at $z<4$, is crucial in determining what causes low-mass galaxies to permanently quench as opposed to resuming cold gas accretion and future star formation.

\section*{Acknowledgements}
This work is based in part on observations made with the NASA/ESA/CSA James Webb Space Telescope and the NASA/ESA Hubble Space Telescope obtained from the Space Telescope Science Institute, which is operated by the Association of Universities for Research in Astronomy, Inc., under NASA contract NAS 5–26555. The data were obtained from the Mikulski Archive for Space Telescopes at the Space Telescope Science Institute, which is operated by the Association of Universities for Research in Astronomy, Inc., under NASA contract NAS 5-03127 for JWST. These observations are associated with programs JWST-GO-1837, JWST-GO-2561, JWST-ERS-1324, JWST-DD-2756, HST-GO-11689, HST-GO-13386, HST-GO/DD-13495, HST-GO-13389, HST-GO-15117, and HST-GO/DD-17231. SEC and KEW gratefully acknowledge funding from JWST-GO-1837 and JWST-GO-2561 and the Alfred P. Sloan Foundation Grant FG-2019-12514. BW and JL acknowledge support from JWST-GO-02561.022-A. RP and DM acknowledge funding from JWST-GO-2561. PD acknowledges support from the NWO grant 016.VIDI.189.162 (``ODIN'') and from the European Commission's and University of Groningen's CO-FUND Rosalind Franklin program. JSD acknowledges the support of the Royal Society through a Royal Society Research Professorship. FC acknowledges support from a UKRI Frontier Research Guarantee Grant (grant reference EP/X021025/1). Support for this work was provided by The Brinson Foundation through a Brinson Prize Fellowship grant. KG and TN acknowledge support from Australian Research Council Laureate Fellowship FL180100060. Some of the data products presented herein were retrieved from the Dawn JWST Archive (DJA). DJA is an initiative of the Cosmic Dawn Center, which is funded by the Danish National Research Foundation under grant No. 140.

\facilities{JWST (NIRCam)}

\software{
\textsc{Astropy} \citep{astropy2013,astropy2018,astropy2022},
\eazy{} \citep{Brammer2008},
\grizli{} \citep[\url{github.com/gbrammer/grizli}]{grizli},
\galfit{} \citep{Peng2002,Peng2010},
\pros{} \citep{prospector2021},
\textsc{FSPS} \citep{fsps2009,fsps2010a,fsps2010b},
\textsc{Python-FSPS} \cite{pythonfsps2014},
\textsc{Aperpy} \citep[\url{github.com/astrowhit/aperpy}]{Weaver2023},
\textsc{Source Extractor} \citep{Bertin1996},
\textsc{SEP} \citep{Barbary2016},
\textsc{extinction} \citep{extinction},
\textsc{SFDMap} \citep[\url{github.com/kbarbary/sfdmap}]{Schlegel1998,Schlafly2011},
\textsc{Pypher} \citep{Boucaud2016},
\textsc{Photutils} \citep{photutils2022},
\textsc{astrodrizzle} \citep{Gonzaga2012},
\textsc{numpy} \citep{numpy2011},
\textsc{matplotlib} \citep{matplotlib2007}
}

\bibliographystyle{aasjournal}
\bibliography{references}
\end{document}